\documentclass{jaa}
\usepackage{epsfig,amsmath,amssymb,graphicx,soul,color,subfigure,lscape,floatrow,caption,natbib}

\newcommand{\aap}{    {\it Astron. Astrophys.}}

\newcommand{\aas}{   {\it Am. Astron. Soc.}}

\newcommand{\apj}{    {\it The Astrophys. J.}}
\newcommand{\apjl}{   {\it The Astrophys. J. Lett.}}
 
\newcommand{\ceab}{   {\it Central European Astrophysical Bulletin}}

\newcommand{\iau}{    {\it IAU Symp.}}
\newcommand{\mnras}{  {\it Mon. Not. R. Astron. Soc.}}

\newcommand{\solphys}{{\it Solar Phys.}}
 
\newcommand{\ssr}{    {\it Space Sci. Rev.}} 
\newcommand{\ppsc}{   {\it Proceedings of the 12th Python in Science Conference}}

\title[Spatial Distribution of Macrospicules]{Non-homogeneous Behaviour of the Spatial Distribution of Macrospicules}
\author[N. Gyenge, S. Bennett and R. Erd\'elyi]{N. Gyenge$^{1,2}$,  S. Bennett$^{2}$ and R. Erd\'elyi$^{1,2}$\\
$^{1}$Heliophysical Observatory, Research Centre for Astronomy and Earth Sciences,\\ Hungarian Academy of Sciences, Debrecen, P.O.Box 30, H-4010, Hungary\\
$^{2}$Solar Physics and Space Plasmas Research Centre (SP2RC)\\University of Sheffield, Hounsfield Road, Hicks Building, Sheffield S3 7RH, UK}
\pubyear{2014}
\volume{xx}
\date{Received 2014; accepted 2015}

\begin{document}
\maketitle

\label{firstpage}

\begin{abstract}
In this paper the longitudinal and latitudinal spatial distribution of macrospicules is examined. We found a statistical relationship between the active longitude determined by sunspot groups and  the longitudinal distribution of macrospicules. This distribution of macrospicules shows an inhomogeneity and non-axysimmetrical behaviour in the time interval from June $2010$ until December $2012$ covered by observations of the Solar Dynamic Observatory (SDO) satellite. The enhanced positions of the activity and its time variation has been calculated. The migration of the longitudinal distribution of macrospicules shows a similar behaviour as that of the sunspot groups.
\end{abstract}


\section{Introduction}
The behaviour of solar non-axisymmetric activity has been studied since the beginning of the last century \citep{Chidambara}. It has been recognised that the distribution of sunspot groups is not uniform, they tend to cluster to a certain heliographic longitude \citep{Bumba,Balthasar,Wilkinson}. These early studies of the active longitudes focused mainly on the distribution of sunspot groups or sunspot relative numbers. From the middle of the 20th century it has been suggested that not just the sunspot groups have a non-homogeneous longitudinal distribution. This inhomogeneity has been found in the case of solar flares \citep{Zhang}, surface magnetic fields \citep{Benevolenskaya}, heliospheric magnetic field \citep{Mursula} and, recently, active longitudes have been observed in coronal streamers \citep{Jing}.

\newpage
Macrospicules (hereinafter: MS) are chromospheric objects observed in H$\alpha$ and He $30.4$ nm \citep{Bohlin,Wang,Murawski,Scullion}. They are explosive jet-like features extending up to, on average, $29$ Mm and velocities up to approximately $110$ km/s \citep{Zaqarashvili}. Their structure reflects the solar atmosphere they move through, they are proposed to have a cool core, surrounded by a hot sheath \citep{Parenti}. They are of particular use in this study, as they are observed from the solar equator to the poles. 

In this paper we study the longitudinal and latitudinal spatial distributions of MS. Furthermore, we will explore the relationship between the sunspot groups, non-axisymmetric behaviour or Active Longitude (hereinafter: AL) and longitudinal distributions of MS.

\section{Observations and Databases}
The MS were observed using the $30.4$ nm spectral window AIA on-board SDO (Solar Dynamic Observatory) \citep{AIAspec}. This takes a $4096 \times 4096$ pixel, full disc, image of the Sun at a cadence of $12$ s. We took typical samples of two hours, twice a month, from June $2010$ until December $2012$. For each image the solar limb was flattened out, making it easier to identify and measure the MS. They are extremely difficult to measure on disk and as such this study concentrates on those occuring at the limb. We record the time at the moment they become visible at the limb and their angular displacement from solar due east. Measuring MS this way we identified 101 examples of MS. The physical dimensions and the heliographic coordinates have been estimated. 

The source of sunspot data that we use to calculate the most enhanced longitude of sunspot groups (AL) is the Debrecen Photoheliographic Data (DPD) sunspot catalogue \citep{DPD}. This database is the continuation of the classic Greenwich Photoheliographic Results (GPR), the source of numerous works in this field. The sunspot catalogue has been used, providing a time sample from $1974$. This data sample contains information about area and position for every sunspot.

\section{Statistical Study of the Latitudinal Distribution of MS}
To study the latitudinal spatial behaviour of MS as a first step, we determine the heliographical latitudes (B). For further analysis, the Carrington latitudes, $B$, have been transformed, into the following system:

\begin{equation}
	\begin{split}
      		\phi=-(B+90^{\circ})/90^{\circ},  B<0 \\
      		\phi=-(B-90^{\circ})/90^{\circ},  B>0
    \end{split}
\end{equation}

The domain of interest of the quantity $\phi$ is $[-1;1]$. The $\phi=0$ point contains the northern and southern poles. The $[0;1]$ sub-domain of $\phi$ represents the northern hemisphere, the ascending $\phi$ values from $0$ to $1$ show the descending latitudes from $90^{\circ}$ to $0^{\circ}$. The southern hemispheres have been considered in the same way.

\begin{figure}
	\centering
		\includegraphics[width=100mm] {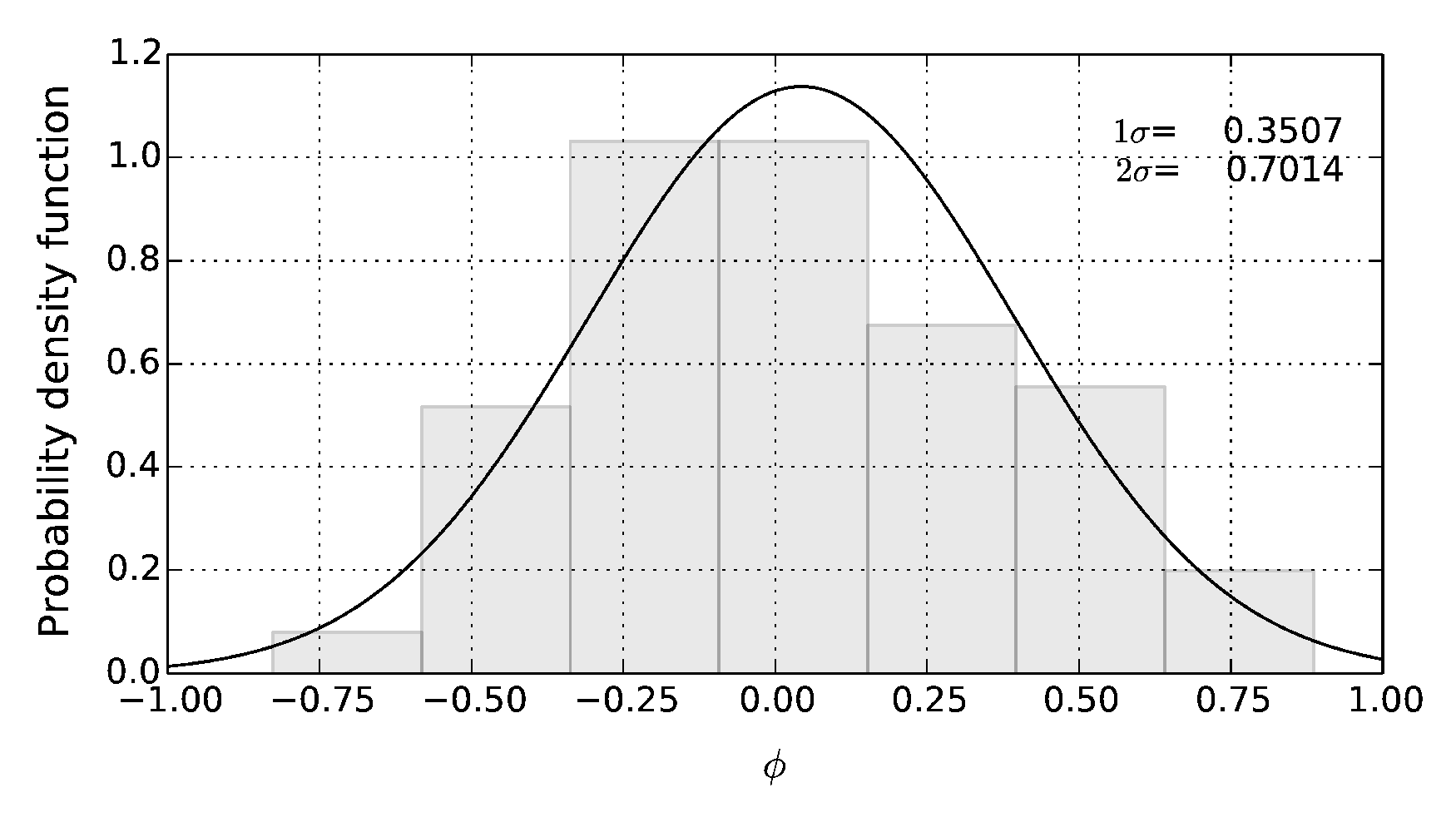}
		\caption{ The gray area shows the probability density function of the parameter $\phi$. The solid black line is the fitted Gaussian distribution. The values standard deviation $1\sigma$ and $2\sigma$ of the normal distribution have been indicated in the top right corner.}
		\label{ms_dis}
\end{figure}

Figure~\ref{ms_dis} shows the result of the statistics above. The histogram depicts a normal distribution. The mean of the distribution is $\overline{\phi}=0.043$, suggesting that most of the MS tend to cluster to the poles. We also found that the northern hemisphere was a slightly more active in this time period. The standard deviation $1\sigma=0.3507$ and $2\sigma=0.7014$. Hence, $68\%$ of the data tend to cluster in a $31.5^{\circ}$ wide belt from the poles. That is to say: $68\%$ of MS are between the $\pm58.5^{\circ}$ and $\pm90^{\circ}$ heliographic latitude, and, $95\%$ of MS are in a $63^{\circ}$ degrees belt from the poles or between the $\pm27^{\circ}$ and $\pm90^{\circ}$ in heliographic latitude. Therefore, MS are able to exercise longitudinal inhomogeneity at higher latitudes.

\section{Statistical study of longitudinal distribution of MS}
\subsection{Activity maps of active longitudes based on sunspots}

\begin{figure}
	\centering
		\includegraphics[width=128mm]{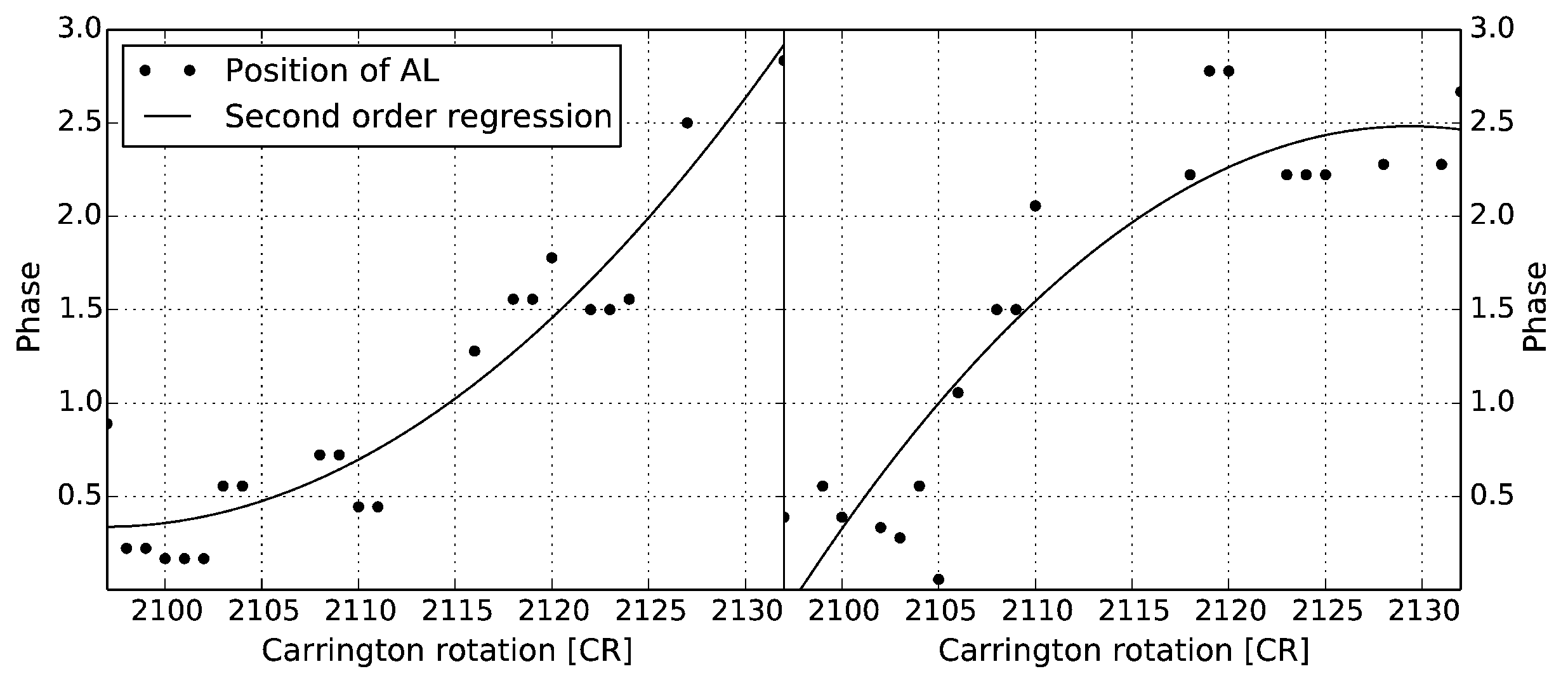}
		\caption{The migration of the active longitudes in the time interval of CR $2097$ to $2128$ based on sunspot groups. The left panel sows the northern hemisphere. The right panel is the southern hemisphere.}
		\label{AL}
\end{figure}

According to our previous study \citep{Gyenge2014} the active longitudes' identification method have been considered and the active longitude was found to be distinct in each hemisphere. The present investigation started with a similar method as described in our preceding paper \citep{Gyenge2012}. The areas and positions of all sunspot groups are considered. The solar surface is divided into longitudinal bins of $20^{\circ}$ and the areas of all groups were summed up in each bin: $ A_{i}$ in certain Carrington Rotation (CR) between $2097$ and $2128$, which is the time interval of the MS sample. Next, the longitudinal activity concentration is represented by the quantity $W$ defined by,
 
\begin{equation}
      W_{i,CR} = \frac{A_{i,CR}}{ \sum_{j=1}^{N} A_{j,CR} },
\end{equation}

where $N$ is the number of bins, $\sum_{j=1}^{N} A_{j,CR}$  is the sum of all sunspot groups in a given CR and $A_{i,CR}$ is the  total area of sunspot groups in a CR and at a specific longitudinal bin.

In each CR we omitted all of the $ W_{i,CR}$ values which are lower than the $3\sigma$ significance limit. The highest peak, $AL_{CR}$, has been selected from this decayed sample (which contains only the significant peaks) caused by the significance test. For further analysis, the Carrington longitudes, $\lambda$, will now be transformed, into Carrington phase period: 

\begin{equation}
	\psi = \lambda/360^{\circ}.
\end{equation}

Hence, the values of the phases are always smaller or equal (which is the entire circumference) than one. 

The time-variation of the parameter $AL_{CR}$ is plotted in Figure~\ref{AL}. The vertical axis is the phase parameter, which has been repeated by $3$ times.
The northern (left-hand-side) and the southern (right-hand-side) cases are considered separately. Both figures unveil a clear increasing migration pattern. \cite{Usoskin,Gyenge2014} found similar patterns at a different time interval. Most of the migration follows a parabola shape (which has been fitted by the least-square method).

\subsection{Relationship between the AL and MS longitudinal distribution}

\begin{figure}
	{\includegraphics[width=90mm]{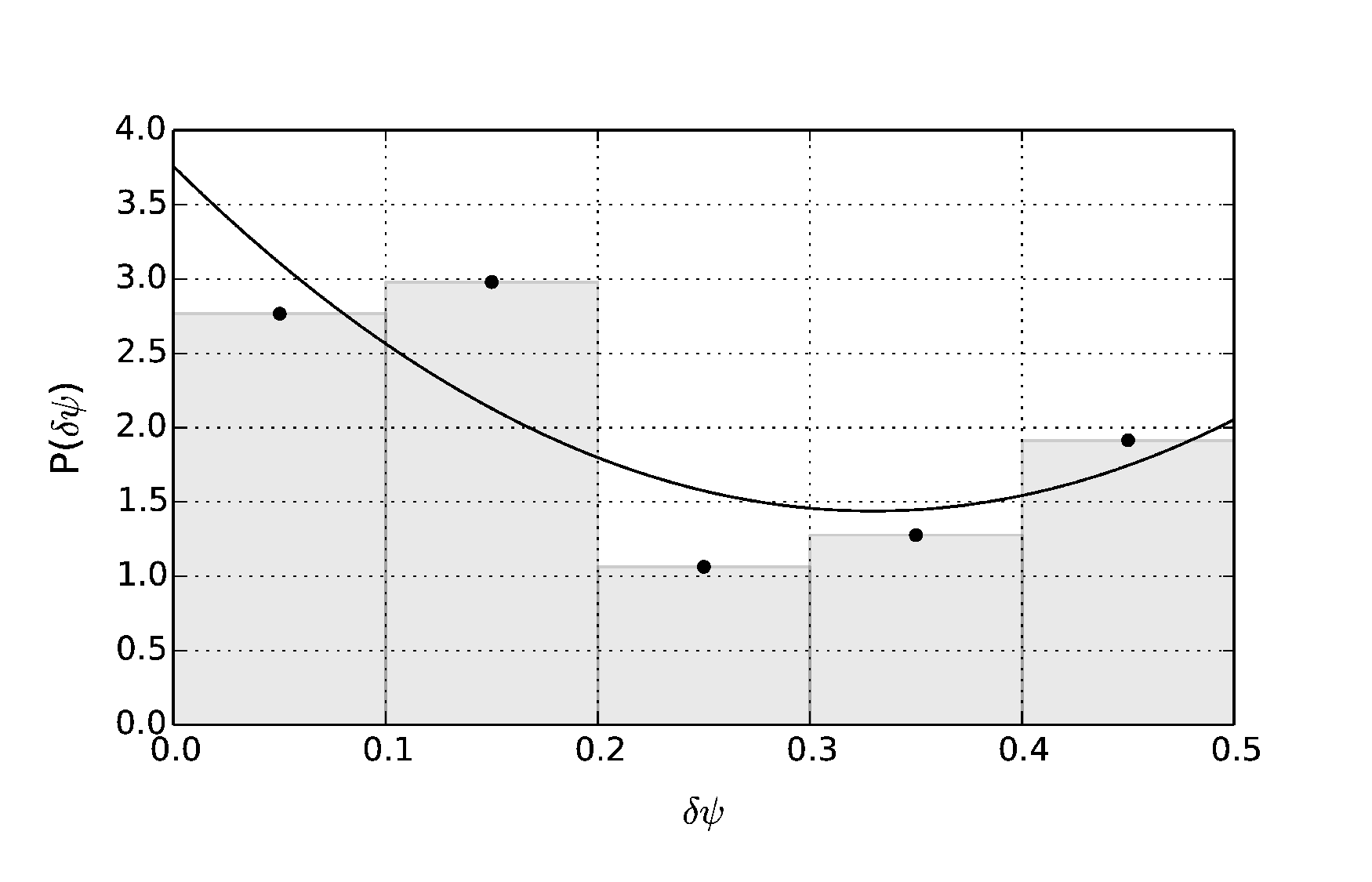}}
	{\caption{The probability density function (PDF) of the $\delta\psi$ parameter. }\label{stat}}
\end{figure}	

The parameter $\delta\psi$ is now introduced to study the relationship between the active longitude  $AL_{CR}$ defined by sunspot groups, and the longitudinal position of MS $L_{CR}$ in CR. 

\begin{equation}
      \delta\psi = \left| AL_{CR} - L_{CR}\right|.
\end{equation}

The parameter $\delta\psi$ has been reduced by a unit phase if it is larger than $0.5$, which means this quantity represents the shortest phase difference between the longitudinal position of a given MS and the position of active longitudes in both hemispheres. For further analysis, the $\delta\psi$ samples of the northern and southern hemispheres are now combined. 

The probability density function (PDF) of the quantity $\delta\psi$ is shown in Figure~\ref{stat}. On the $x$-axis the meaning of the lower values reflect on the smallest longitudinal difference in phase, the value $0.5$ phase jumps to the opposite side of the Sun.

The MS tend to cluster near the active longitudes, which is shown by the first and second peaks: $\delta\psi< 0.2 (<\pm 36^{\circ}$) 61 \% of the candidates. However, there is a significant peak around $0.5$, which is the signature of the appearance of secondary longitudinal belts. The secondary belt exists always at the same time as the primary. Note that the latter is always stronger than the secondary belt, and the phase shift is around $0.5$. The MS show a similar behaviour. A secondary belt appears for the $22\%$ of the events and $\delta\psi< 0.1 (<\pm 18^{\circ}$).

\section{Results and Discussion}
We investigated the distribution of macrospicules detected at the solar surface as function of their longitudinal and latitudinal coordinates in Carrington coordinates.

A non-homogeneous latitudinal macrospicule distribution has been found. Most of the events tend to cluster to the higher latitudes ($95\%$ of MS are with in the $\pm27^{\circ}$ to $\pm90^{\circ}$ heliographical latitude). The number of the events is found to be growing exponentially from the equator to the pole in both hemispheres.

A slightly asymmetrical behaviour has been found between the two hemispheres in the studied time interval, where the northern hemisphere was marginally more active than the southern. In the studied time period other phenomena show northern-southern asymmetry. The northern hemisphere was significantly more active in terms of the averaged maximum sunspot area (northern hemisphere: 203, southern hemisphere: 183) and the international sunspot number (northern hemisphere: 29, southern hemisphere: 18). For that reason we assume the location of the MS might be connected to the solar dynamo processes, however, caution must be exercised as this conclusion is based on a rather limited statistical sample.

The longitudinal spatial distribution of MS is not uniform either. A large proportion of the MS (83 of from the 101 in our sample) tend to cluster to the AL. In the case of the primary active longitude belt, the macrospicules are within $\pm 36^{\circ}$ degrees of the active longitude. The secondly belt has a $\pm 18^{\circ}$ wide range where the macrospicule are found to be concentrated. This supports the existence of an active longitude at higher latitudes. MS instances extend up to 50 Mm in the solar atmosphere. Therefore, they have been proposed as a viable candidate for solar wind acceleration and coronal heating, see e.g. \cite{Pike}.

The origin of such MS is either due to wave modes (e.g., \cite{Zaqarashvili, Scullion}) or by magnetic reconnection \citep{Wilhelm, Heggland, Murawski}. However, how these triggering mechanisms of MS are related with magnetic flux transport processes is not yet known. Recently, \cite{Kayshap} reported the evolution of a small-scale bipolar flux tube in quiet-Sun and its internal reconnection to produce MS.  This provides some clues on the connection of localised generation of macrospicule with small-scale magnetic field evolution. However, more statistics in terms of the observations and stringent modelling are needed to explore this aspect of MS origin.

The temporal variation of the number of MS has been investigated, but the correlation between the temporal density of MS and the increasing trend of Solar Cycle 25 has not yet been found. A reason for the lack of finding such correlation could be the limited time period and sample size of the data investigated.

A large sample and more comprehensive statistical study is now in preparation for a more detailed search for further identifiable non-homogenous longitudinal distributions of MS in the entire time period covered by observations of the SDO satellite.

\section*{Acknowledgements}
We would like to thank for the invitation, support and hospitality received from the Hungarian Academy of Sciences under their Distinguished Guest Scientists Fellowship Programme (ref. nr. 1751/44/2014/KIF ) that has allowed him to stay three months at the Debrecen Heliophysical Observatory (DHO) of the Research Centre for Astronomy and Earth Sciences, Hungarian Academy of Sciences during which this research was undertaken. SB acknowledges the warm hospitality received from DHO while visiting there and carrying out this research. The research leading to these results has also received funding from the European Community's Seventh Framework Programme (FP7/2007-2013) under grant agreement eHEROES (project No. 284461) and from Science and Technology Facilities Council (STFC), UK. This research was made use of SunPy, an open-source and free community-developed Python solar data analysis package \citep{Mumford}.

\end{document}